# Explainable Facial Expression Recognition for People with Intellectual Disabilities

Silvia Ramis
Group of Computer Graphics, Computer Vision and IA
Maths and Computer Science Department
Universitat de les Illes Balears
Palma, Illes Balears, Spain
silvia.ramis@uib.es

Cristina Manresa-Yee
Group of Computer Graphics, Computer Vision and IA
Maths and Computer Science Department
Universitat de les Illes Balears
Palma, Illes Balears, Spain
cristina.manresa@uib.es

Jose M. Buades-Rubio
Group of Computer Graphics, Computer Vision and IA
Maths and Computer Science Department
Universitat de les Illes Balears
Palma, Illes Balears, Spain
josemaria.buades@uib.es

F. Xavier Gaya-Morey
Group of Computer Graphics, Computer Vision and IA
Maths and Computer Science Department
Universitat de les Illes Balears
Palma, Illes Balears, Spain
francesc-xavier.gaya@uib.es

## ABSTRACT

Facial expression recognition plays an important role in human behaviour, communication, and interaction. Recent neural networks have demonstrated to perform well at its automatic recognition, with different explainability techniques available to make them more transparent. In this work, we propose a facial expression recognition study for people with intellectual disabilities that would be integrated into a social robot. We train two well-known neural networks with five databases of facial expressions and test them with two databases containing people with and without intellectual disabilities. Finally, we study in which regions the models focus to perceive a particular expression using two different explainability techniques: LIME and RISE, assessing the differences when used on images containing disabled and non-disabled people.

## KEYWORDS

Facial Expression Recognition, Social Robots, Intellectual Disabilities, Explainable Artificial Intelligence, Human-Computer Interaction.

## 1 Introduction

The study of Facial Expressions Recognition (FER) is a very extended field in the area of artificial vision [1]. Computer vision makes it possible to acquire, process, analyse and understand images taken by one or more cameras. Often, computer vision and Human-Computer Interaction (HCI) [2] or Human-Robot Interaction (HRI) [3] go hand in hand. HRI is a relatively new field compared to HCI and, therefore, many techniques used in HCI are also used in HRI [3].

A social robot must be able to express and/or to recognize emotions, communicate through dialogue, use natural gestures, have a personality, and establish social relationships. Studies such as [4] affirm that humans prefer to interact with machines in the same way that they interact with other people. These robots can be used as toys, educational tools, or even for therapeutic aids [4]. They have also been shown to improve learning skills in adults with intellectual disabilities. In [5] they explore how interactions with a social robot can contribute to the learning of adults with intellectual disabilities and how they want to interact with these robots, where the results suggested that both the physical presence of the Pepper robot and the support of the tablet play an important role in engaging adults with intellectual disabilities. On the other hand, in [6] they explore the impact of the use of social robots in a prolonged way (for 24 months) in the educational context with secondary school students with intellectual disabilities and autism, where the results showed that the participants responded positively to the use of robots in the school and would recommend it to other schools.

For a social robot to be able to communicate, have a personality and establish relationships, it must be able to express and/or recognize emotions. Neural networks are used to study the recognition of facial expressions [7, 8, 9, 10]. Despite the good performance of these works, many times we do not understand why an expression that seems correct to the human eye is misclassified. To cope with this problem, there are explainability techniques that serve to provide more information about the inner workings of a neural network and make it more transparent. In this way, we can understand, for example, the most important facial regions in the classification of facial expressions [11] and compare them to the most important regions for the human eye. On the other hand, normally, these systems are trained and tested with people without

*Article Title Footnote needs to be captured as Title Note
†Author Footnote to be captured as Author Note

disabilities. But are these systems viable for working with people with disabilities?

This work studies the development of a FER system addressed to people with intellectual disabilities that would be integrated into a social robot. For this, we pose two main questions:

**1.** Can existing neural networks trained with facial expressions predict the facial expressions of people with intellectual disabilities? (Q1).

**2.** Are there differences between the facial expressions of people with and without intellectual disabilities? (Q2).

First, two neural networks (AlexNet and VGG19) are trained with five known databases of facial expressions (CK+, JAFFE, BU4-DFE, WSEFEP and FEGA), which have been used for cross-database experimentation [12], and then the models are tested with two databases: one containing people with intellectual disabilities (MuDERI) and another without them (FE-test).

Secondly, the predictions of the models over the two datasets are explained through the use of two eXplainable Artificial Intelligence (XAI) techniques: LIME and RISE, to understand the differences due to human diversity.

The work is structured as follows: in the following section, a review of the related literature is carried out to identify the most relevant works related to the topic. Section 3 describes the datasets, image pre-processing techniques, Convolutional Neural Networks (CNNs), and the XAI techniques used. Sections 4 and 5 present the experimentation and the obtained results, and these are discussed in Section 6. Finally, the last Section concludes the work and summarizes the main contributions.

## 2 State of the Art

Emotions, which are externalized through facial expressions, play an important role in human behaviour, communication, and interaction. Therefore, the recognition of these expressions in an automatic way is of high interest and plays an important role in HCI.

Works as [7, 8, 9, 10] studied FER to integrate it into HCI applications [7, 8] or use it in social robots [9,10]. In [7] they used previously trained networks for the recognition of facial expressions. For this, they used the CK+ database and adapted four neural networks for the recognition of seven basic expressions, resulting in an average accuracy of 96%. In [8] they used several deep learning algorithms to monitor the users' behaviour and emotions through facial expression and gaze recognition captured by a webcam. The idea was to allow for "in the wild" data collection during user interactions with a web application/platform simply by using a normal webcam. For FER, they trained with CK+, FER+, and AffectNet, and tested with EmotioNet, achieving an accuracy of 75.48%. In the case of the gaze, 54 participants were used for the training set and 20 participants for the test set. In [9], they intended to develop an unmanned flying social robot to monitor dependent people at home, detect the person's condition and provide the necessary assistance. To do this, they used a face detection algorithm and a CNN with a performance of 85% to detect the face and classify it into one of the 7 basic expressions (surprise, fear, happy, sad, disgust, angry, or neutral expression). In [10], the authors used a social robot to evaluate a neural network trained on facial expressions in a real environment. To do this, they carried out a comparison between the accuracy of a CNN and 10 human experts, in addition to analysing the interaction, attention and difficulty in making a certain expression by 29 non-expert users. The results showed that the CNN was 13% less accurate than the experts.

Even though neural networks obtain good results in FER, they are still considered black boxes that need to be understood to verify their proper functioning. A clear example can be found in [13], where the classification between husky dogs and wolves using a neural network is carried out. To find out how the network behaved, they used LIME, a XAI technique based on Local Interpretable Model-agnostic Explanations. The results showed that the network did not look at the dog's or the wolf's features, but rather at the background of the image, using the presence or absence of snow as a key factor in the classification.

Recently, to provide more insight into the inner workings of a neural network and make it more transparent, we found works that apply these techniques in FER to understand automatic emotional annotation [14] or to understand facial regions influential in classification [11]. Heimerl et al. [14] included XAI techniques in their emotional behaviour annotation tool aimed at non-expert users. Participants labelled four of Ekman's six basic facial expressions (happy, sad, angry, and disgust) aided by confidence values of the predicted annotation, as well as visual explanations using XAI (LIME [15], INNvestigate [16]). Weitz et al. [11] trained a CNN to distinguish facial expressions of pain, happy, and disgust. They applied two XAI methods: LRP and LIME, and they observed that the CNN was not focusing exclusively on the face, but also on the background of the image.

With all these in mind, in this work, a FER system is constructed using two different neural networks, and explainability techniques are employed to better understand the models' classification, especially on images containing people with intellectual disabilities.

## 3 Methodology

In this Section, we explain the datasets, pre-processing and CNNs used, in addition to the explainability techniques used to understand the performance of the models.

### 3.1 Datasets

We have used up to seven datasets in this study. The first four are standard datasets widely used in FER studies: the Extended Cohn-Kanade (CK+) [17], the BU-4DFE [18], the JAFFE [19] and the WSEFEP [20] datasets. The Extended Cohn-Kanade (CK+) dataset [17] contains 593 sequences captured from 123 subjects. Each sequence is labelled with one of the 7 basic facial expressions (angry, contempt, disgust, fear, happy, sad, and surprise). The BU-4DFE dataset [18] contains 606 sequences captured from 101 subjects. Each subject has six sequences, one per each facial expression (angry, disgust, fear, happy, sad, and surprise). The Japanese dataset, JAFFE [19] contains 213 images captured from 10 female Japanese actresses and the WSEFEP [20] dataset contains 210 images captured from 30 individuals. Each image of each one of these datasets is labelled with one out of 7 facial expressions (angry, disgust, fear, happy, sad, surprise and neutral). In addition to these four datasets, other two are used in the study: FEGA and FE-test [5]. FEGA is a dataset labelled with Facial Expression, Gender and Age simultaneously. This dataset contains 51 subjects. Each subject performed the seven basic facial expressions, repeating each one eight times, and taking a snapshot each time. On the other hand, FE-test is a dataset composed of 210 frontal images “in the wild”, where each image was labelled with one out of 7 facial expressions.

Finally, we used the MuDERI dataset [21]. MuDERI is a multimodal database containing 12 participants with intellectual disabilities. It is composed of two audio-visual recordings from each participant. In the first recording, positive stimuli are given to the participant to express positive emotions, while in the second recording, negative stimuli are given to express negative emotions. These videos are divided by several timestamps. Each timestamp is annotated with three basic facial expressions (happy, sad, and angry) by five caregivers [21], in addition to be annotated with EEG signals, EDA signals and Kinect data that were synchronized with the audio-visual recordings using these timestamps.

## 3.2 Image pre-processing

The image pre-processing used in this work was face detection, face alignment, and cropping. First, the face detection was performed using the “a contrario” framework proposed by Lisani et al. [22]. Second, we used the 68 facial landmarks proposed by Sagonas et al. [23] to find the eyes and align the face. To do this, we calculated the geometric centroid of each eye and the distance between them to draw a straight line, compute the rotation angle, and align the eyes horizontally. Finally, we cropped the face and resized it to the size required by the CNNs being used.

## 3.3 Convolutional neural networks

In this work we implemented two well-known CNNs: AlexNet and VGG19 [24, 25] for the FER task. VGG19 was used with pre-trained weights with ImageNet, while AlexNet was used without a pre-training. In both cases we used k-cross-validation [5] with k=5 to compute the accuracy.

The VGG19 architecture was proposed by [25] and contains 14 convolutional layers, 5 max pooling layers and 3 fully connected layers. This architecture uses images of 224x224 pixels as input data and is available in Caffe and PyTorch.

The AlexNet architecture was proposed by [24] and is a relatively simple and well-known neural network containing 5 convolutional layers, 3 max pooling layers and 3 fully connected layers. This architecture uses images of 224x224 pixels as input data and is available in Caffe, Keras and PyTorch.

## 3.4 XAI approach

Explainable artificial intelligence focuses on the understanding of the predictions of increasingly complex models, like the two neural networks employed in the current work. As stated in [26], the objectives of XAI are multiple: to increase the trustworthiness of the models, to find causality between data variables, to make transferability to the society easier, and to support decision making, among others. Since we use different models in this work, we have focused on model-agnostic XAI techniques, which allow to understand the predictions of any model. Furthermore, we focus on local explanations to understand individual predictions of the models, to then approximate global explanations for each model and class.

To obtain the explanations, we have employed two different XAI techniques very extended in the image domain: LIME [27] and RISE [28]. Both methods’ basic functioning consists of the creation of a small dataset containing perturbed versions of the image to be explained (see Figure 1). Then each sample of this dataset is passed through the model, and finally the resulting predictions are used to assign each region a relevance for a specific class.

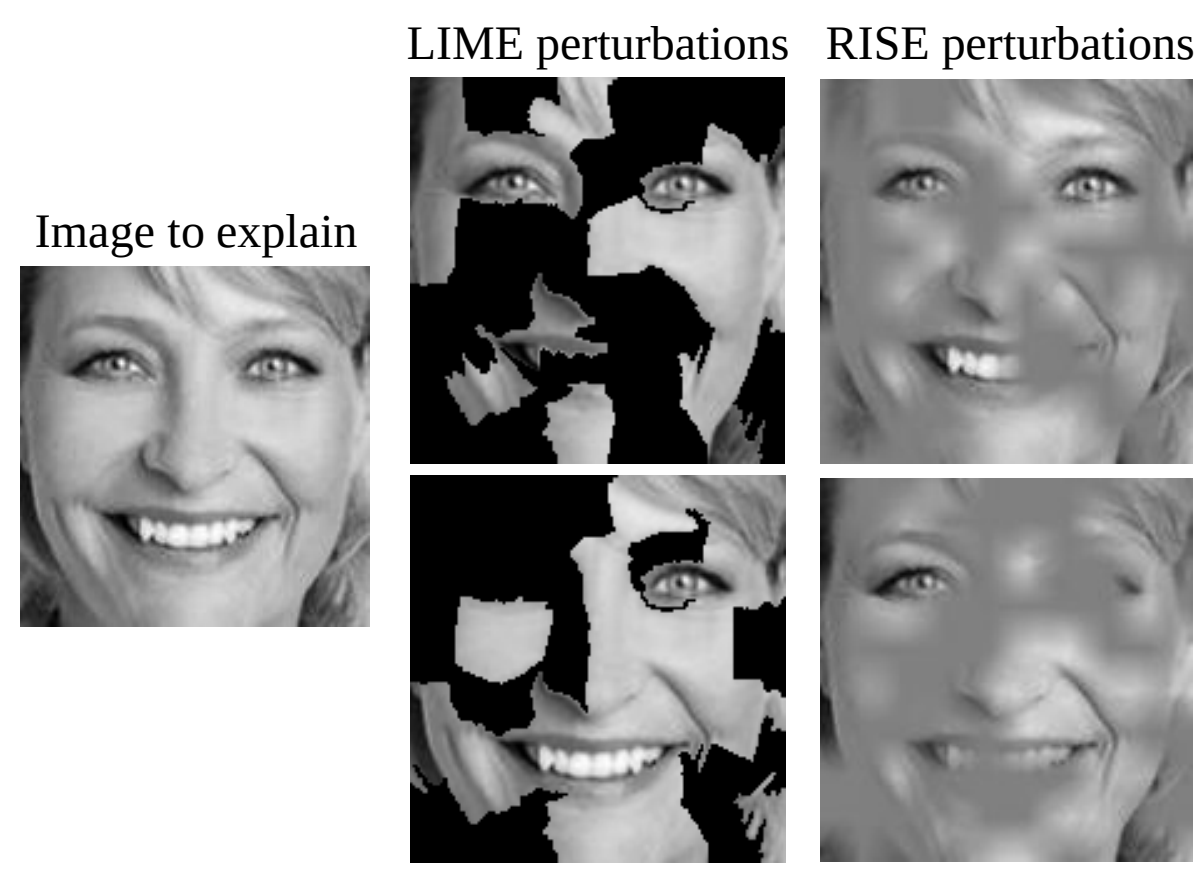

**Figure 1: Perturbed versions of the image being explained, used by LIME (middle column) and RISE (right column).**

In LIME, the regions of the image are usually superpixels, although other segmentation techniques can be used; the occlusion of these regions is done by setting them to a fixed colour, commonly black, although other options can also be used (like using the mean colour value of the regions being occluded); and the relevance of the regions is found out by training a simple, understandable additive model. In our experiment, we have used SLIC [29] to compute the superpixels, setting the number of regions to be approximately 30; we have set the colour of the occlusions to black; and we have centred only on the positive relevance of the regions, omitting negative relevance.

RISE, on the other hand, uses a 2D grid segmentation, and the perturbation is done by upscaling a small grid-like mask with occluded regions in grey to the image size using bilinear interpolation, which allows to avoid hard edges between regions.

Then the relevance of a region is found by computing the average prediction of the model in the samples where it is not masked out.

## 4 Experiments

We have performed three experiments to answer our questions:

### 4.1 Experiment 1

In this first experiment, we used two existing networks, AlexNet and VGG19, to learn the happy, sad, and angry expressions from the combination of several databases (BU4FDE, CK+, JAFFE, FEGA and WSEFEP). Then, we tested them with another database (FE-Test). Neither of these datasets contain participants with disabilities.

This experiment allows to verify that the training is correct, to subsequently test it with a dataset consisting of participants with disabilities.

### 4.2 Experiment 2

To answer the first question raised in the Section 1 (Q1), we test the trained networks from the first experiment on the Muderi dataset [21].

### 4.3 Experiment 3

To answer the second question raised in the Section 1 (Q2), we apply the two XAI techniques (LIME and RISE) explained in Subsection 3.4. The results of the application of both methods can be visualized as a heatmap, where the hottest regions represent the most relevant regions for the prediction of a class by the model. With the aim of summarizing which are the most important regions of the face for a model to predict each class, we have followed the next steps (see Figure 2). First, we select a positives subset from the Muderi dataset and FE-Test of each class, for each model. These will be the examples to be explained locally, and then used to average global explanations. Second, we run LIME and RISE explanations on these positives, storing them in grey scale. Third, we find the face landmarks of each positive and use them to transform the explanations to a normalized space, where the landmarks are always found in the same locations. This will allow the explanations to be invariant to translation, rotation, and scale. Finally, average the normalized explanations for each model, dataset, and class, resulting in global explanations.

## 5 Results

In this Section, we analyse the results of the experiments proposed in Section 4.

### 5.1 Experiment 1

In this first experiment, we train and test with datasets of people without intellectual disabilities using a cross-dataset approach, in order to verify that the models perform correctly on people without intellectual disabilities.

Figure 3 shows the matrix confusion of AlexNet and VGG19 tested on the FE-Test dataset. Although both networks get good classification results, they seem to confuse Sad and Angry in some images, while "Happy" is always predicted correctly.

The accuracy of this experiment is about 86% using AlexNet and about 89% using VGG19, which demonstrates a very good accuracy on images not containing people with intellectual disabilities.

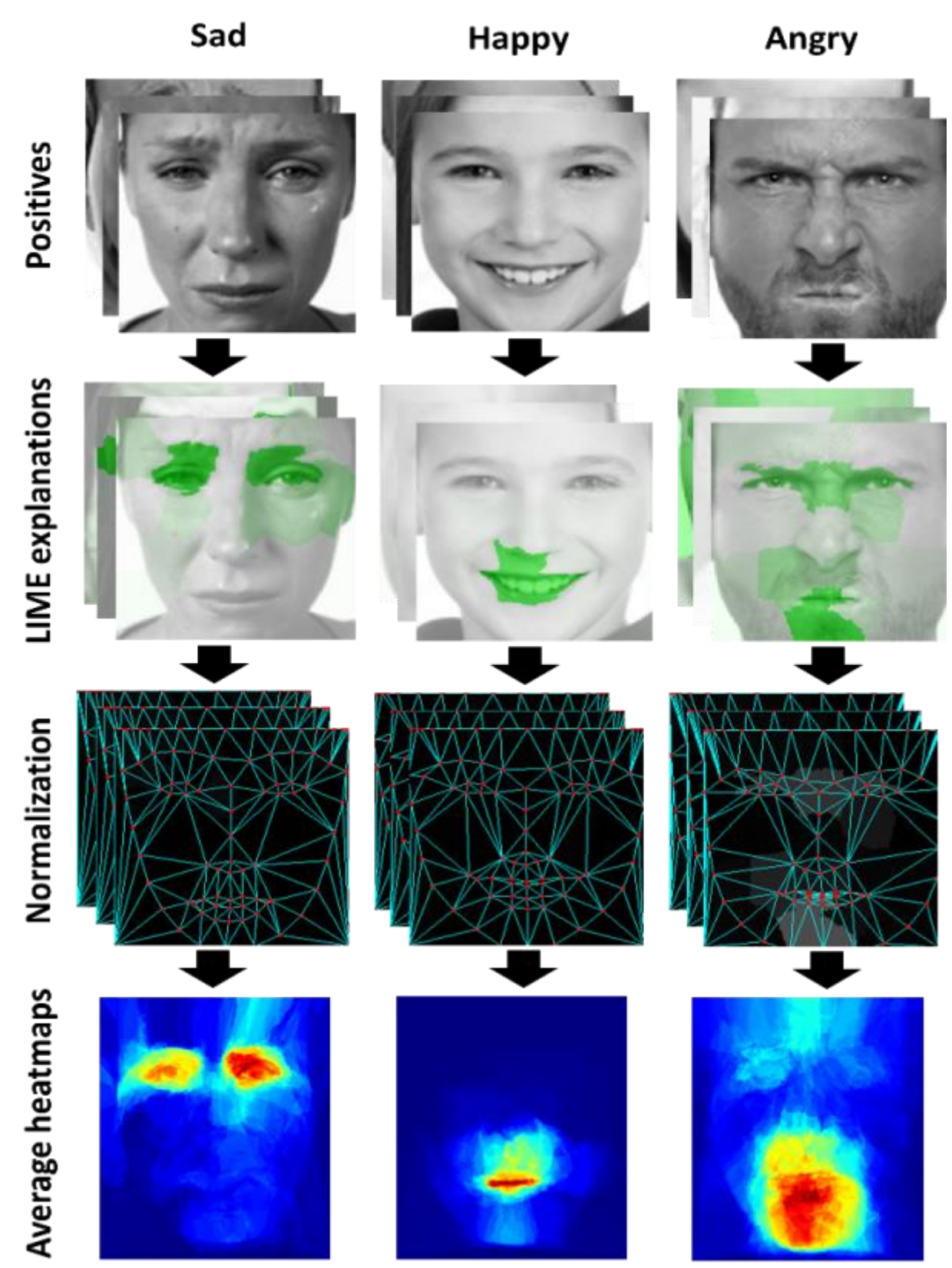

**Figure 2: Summarization process of the most important facial regions for the prediction of each class. By columns, left to right, each of the classes: sad, happy and angry. By rows, top to bottom, all steps followed: selection of positives, explanation using LIME and RISE, transformation to a normalized space using the face landmarks, and average of the results to build heatmaps. Only explanations using LIME for the VGG19 model, and the FE-Test dataset are shown for the sake of simplicity.**

### 5.2 Experiment 2

In this second experiment, we use the models trained in the first experiment and test them with a facial expression dataset of people with intellectual disabilities.

Figure 4 shows the confusion matrices for both networks on the Muderi dataset. Both VGG19 and AlexNet achieve low results in the classification of the three facial expressions (Sad, Happy and Angry) on people with intellectual disabilities. VGG19 demonstrates a tendency towards the Angry expression, over the other two, with 85% of the predicted labels falling in this class. AlexNet, on the other hand, although also showing this tendency towards the Angry expression, gets half of the "Happy" expressions right, meaning that even though not labelling systematically all images with the "Angry" expression, it is challenging to distinguish it from the "Happy" expression.

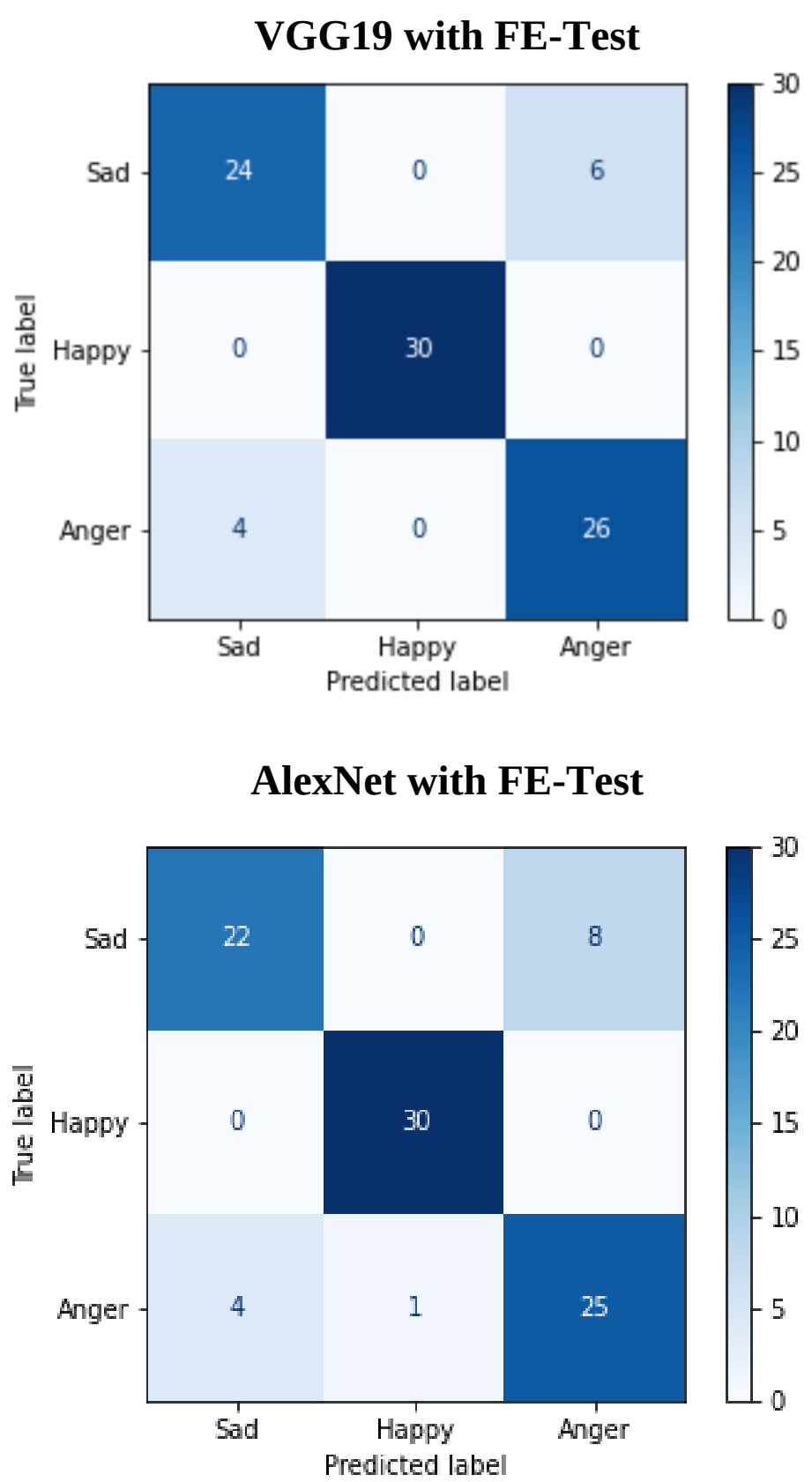

**Figure 3: Confusion matrices for the first experiment: testing with FE-Test dataset, without people with intellectual disabilities.**

The accuracies of the two first experiments are shown in Figure 5, comparing the test results between the Fe-Test and Muderi datasets. As shown, AlexNet gets an accuracy of about 46% on Muderi, while VGG19 only about 19%. This indicates that the networks are not capable of correctly identifying the facial expressions for the Muderi dataset, while they are for the Fe-Test dataset, which leads us to think that people with and without intellectual disabilities do not share the same facial expressions.

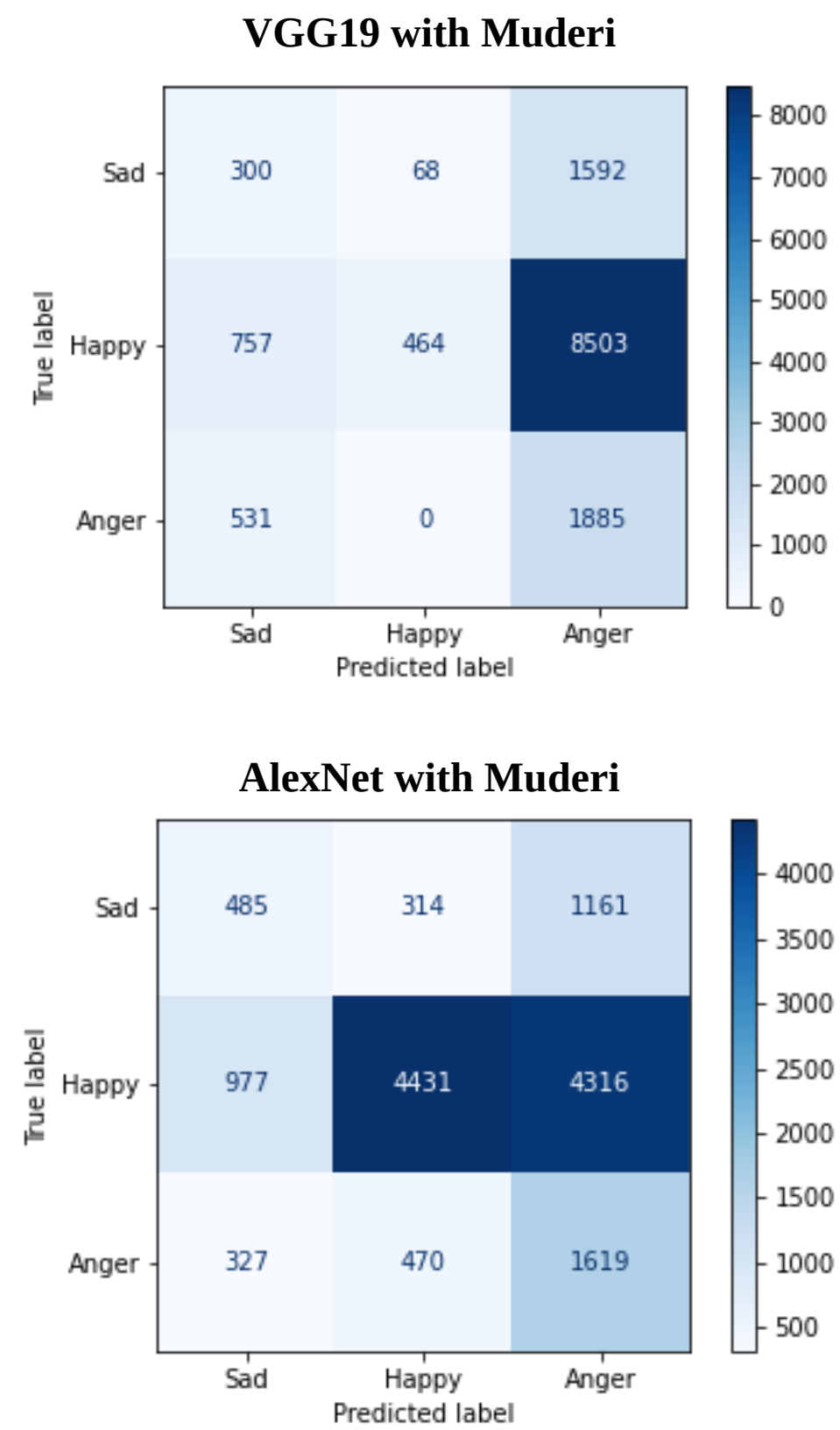

**Figure 4: Confusion matrices for the second experiment: testing with Muderi dataset, containing people with intellectual disabilities.**

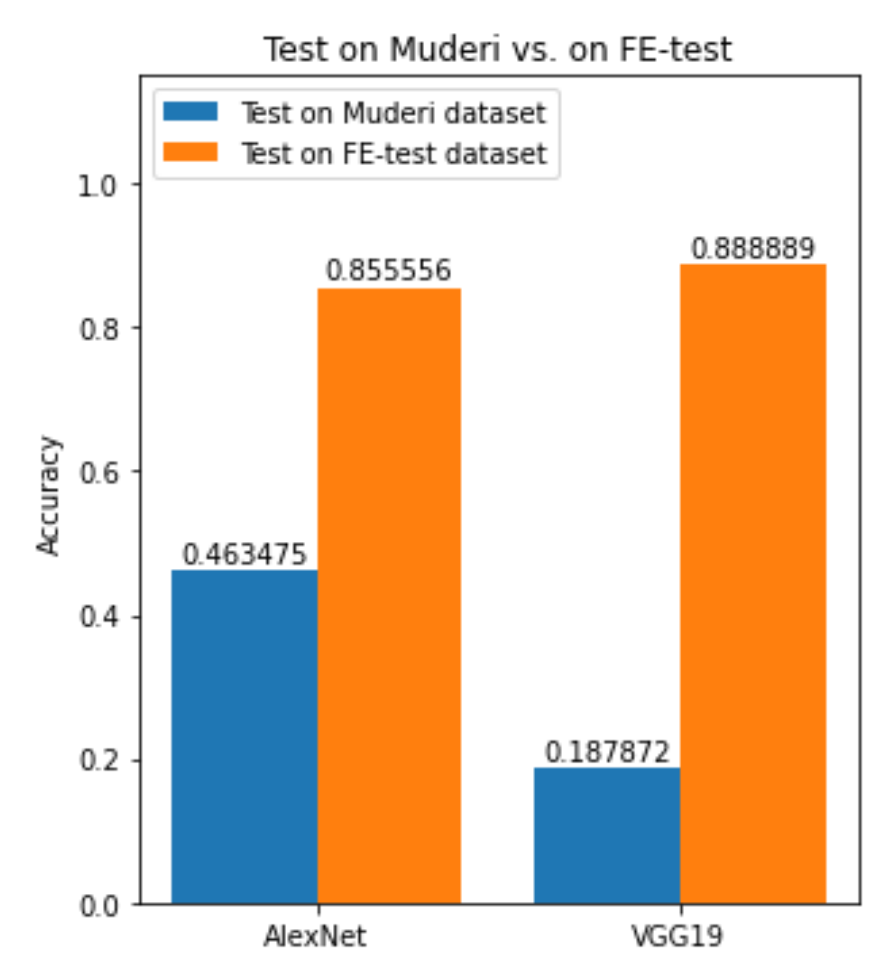

**Figure 5: Accuracies of the test on Muderi and Fe-Test datasets using VGG19 and AlexNet.**

## 5.3 Experiment 3

The results of this experiment are a set of heatmaps identifying the most important regions of the face for the recognition of each facial expression. We have computed a heatmap for each of the two networks (AlexNet and VGG19), with each test dataset (FE-Test and Muderi), using two XAI techniques (LIME and RISE), for each expression (Happy, Sad and Angry), ending up with a total of 24 heatmaps. These heatmaps are shown in Figures 6, 7, 8 and 9.
In following subsections, the differences between these heatmaps are assessed.

#### *5.3.1 Differences between networks and between facial expressions*

From the heatmaps shown (Figures 6, 7 and 8), the greater differences can be found between models and facial expressions, rather than between datasets or XAI techniques. More precisely, AlexNet and VGG19 seem to focus on different parts of the face, even for the same facial expression: for "Sad", VGG19 focuses mainly on the eyes, while AlexNet seems to focus more on the mouth and nose; for "Happy", both nets seem to agree on the important region: mainly the mouth (the smile, usually showing teeth); and for "Angry", VGG19 focuses both on the mouth, eyes, and forehead, while AlexNet focuses only on the area around the eyes and on the forehead.

#### *5.3.2 Differences between XAI techniques*

There are few differences between LIME and RISE, which seem to indicate high reliability of the explanations. In general, LIME heatmaps are more precise, while RISE heatmaps are more blurred, due to their internal functioning. But they seem to coincide in which are the important regions for the models. Perhaps one exception is found on Muderi for the "Sad" expression, where LIME heatmaps focus on the forehead and the area around the nose (Figure 7), while RISE's focus on the eyes (Figure 9).

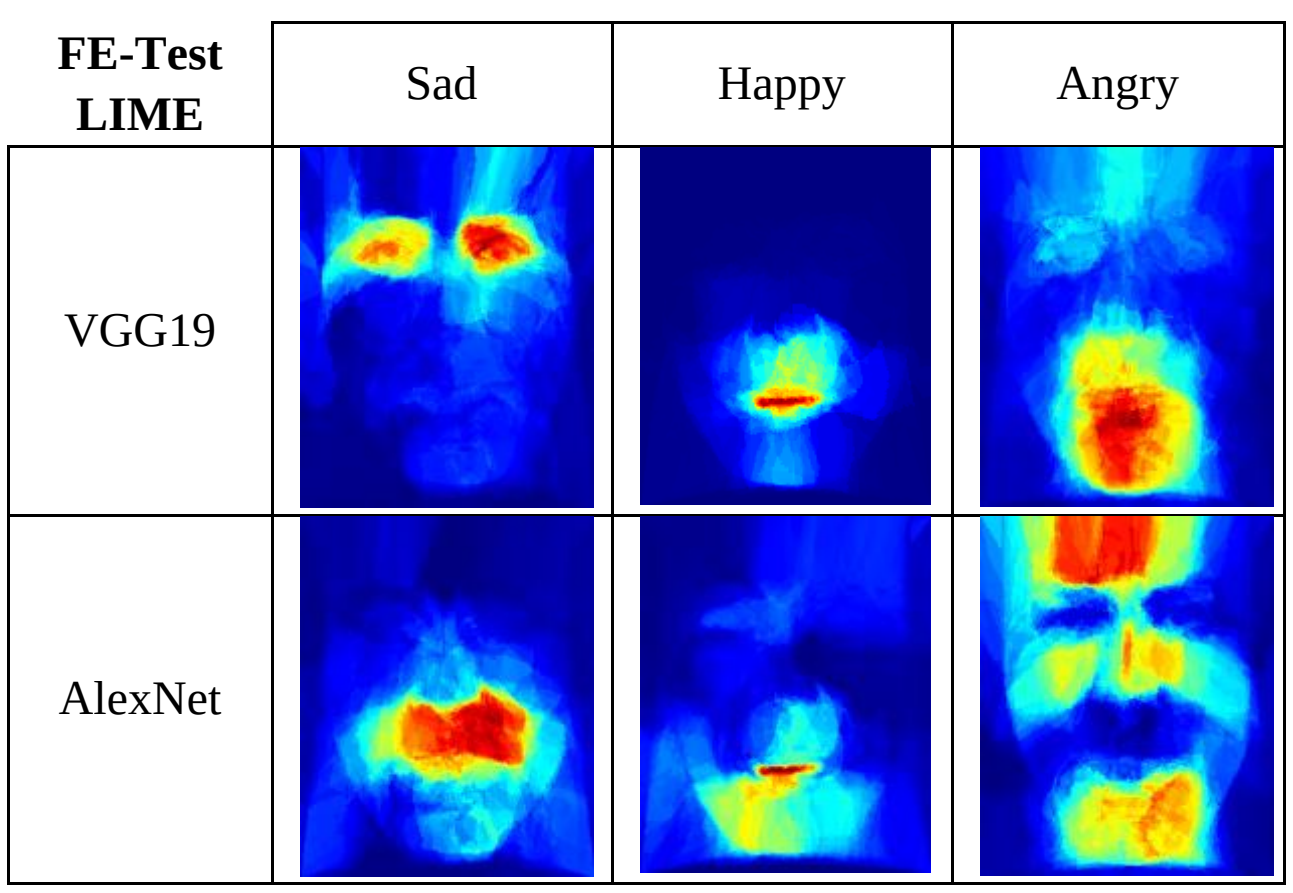

**Figure 6: Results of Lime on FE-Test dataset using both VGG19 and AlexNet neural networks.**

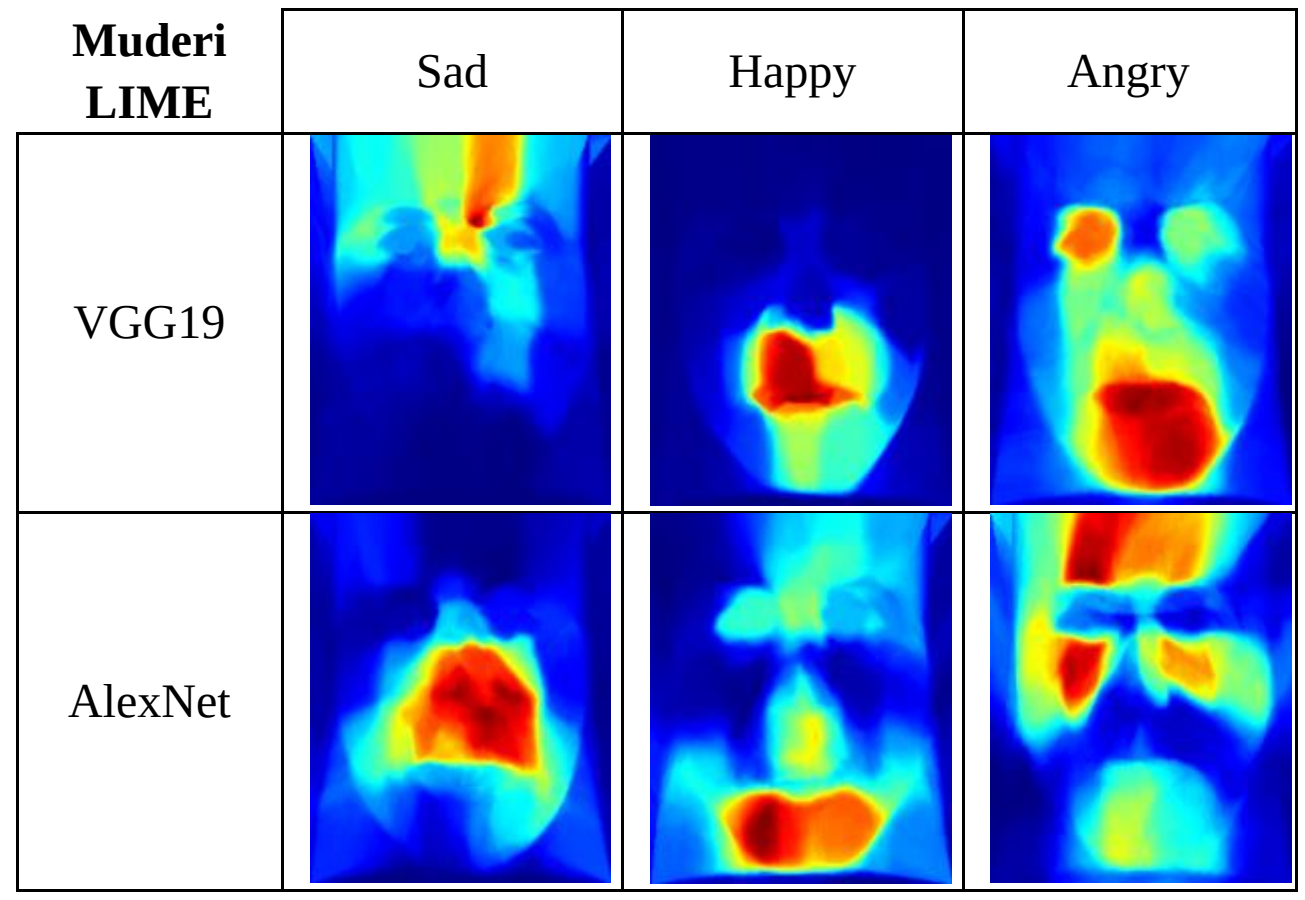

**Figure 7: Results of Lime on Muderi dataset using both VGG19 and AlexNet neural networks.**

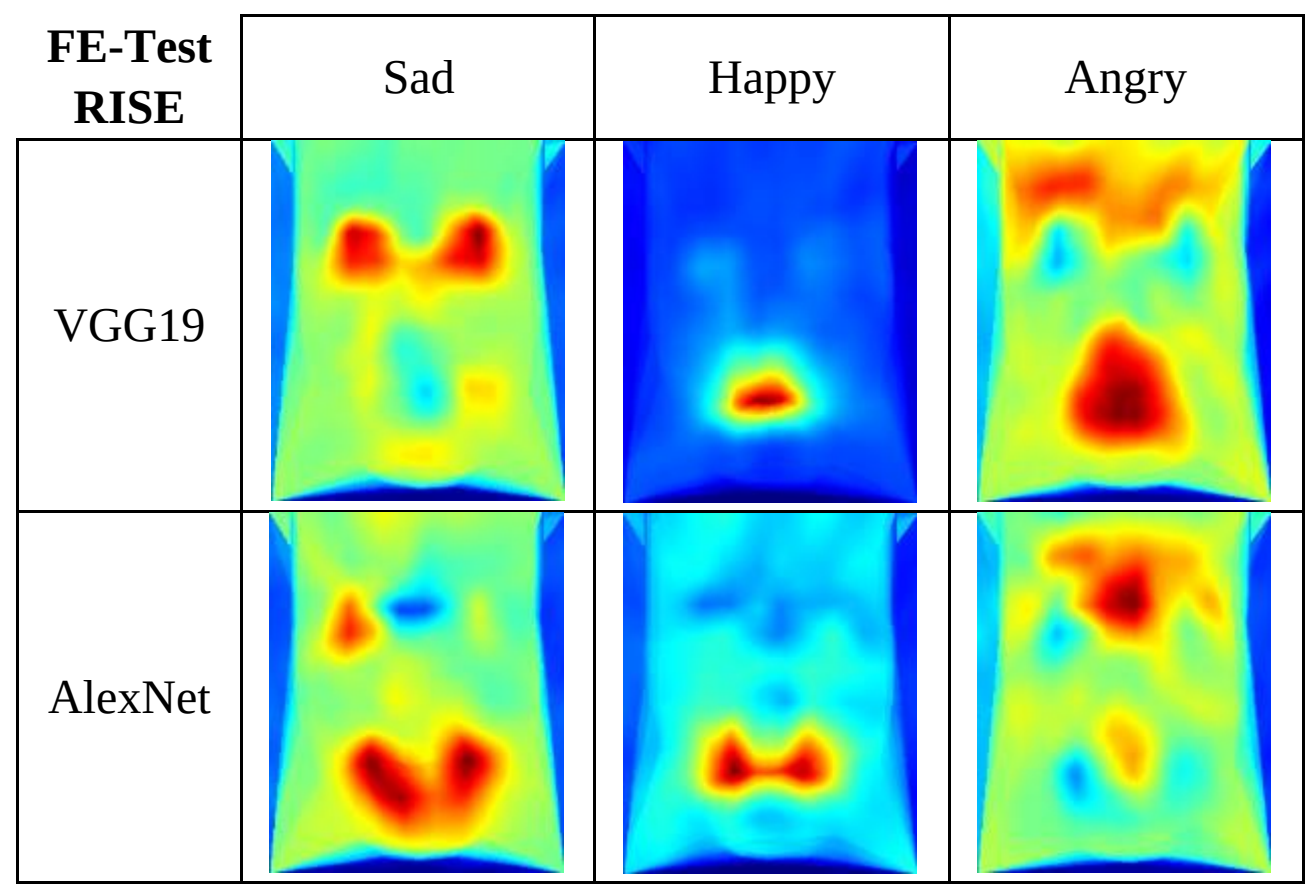

**Figure 8: Results of Rise on FE-Test dataset using both VGG19 and AlexNet neural networks.**

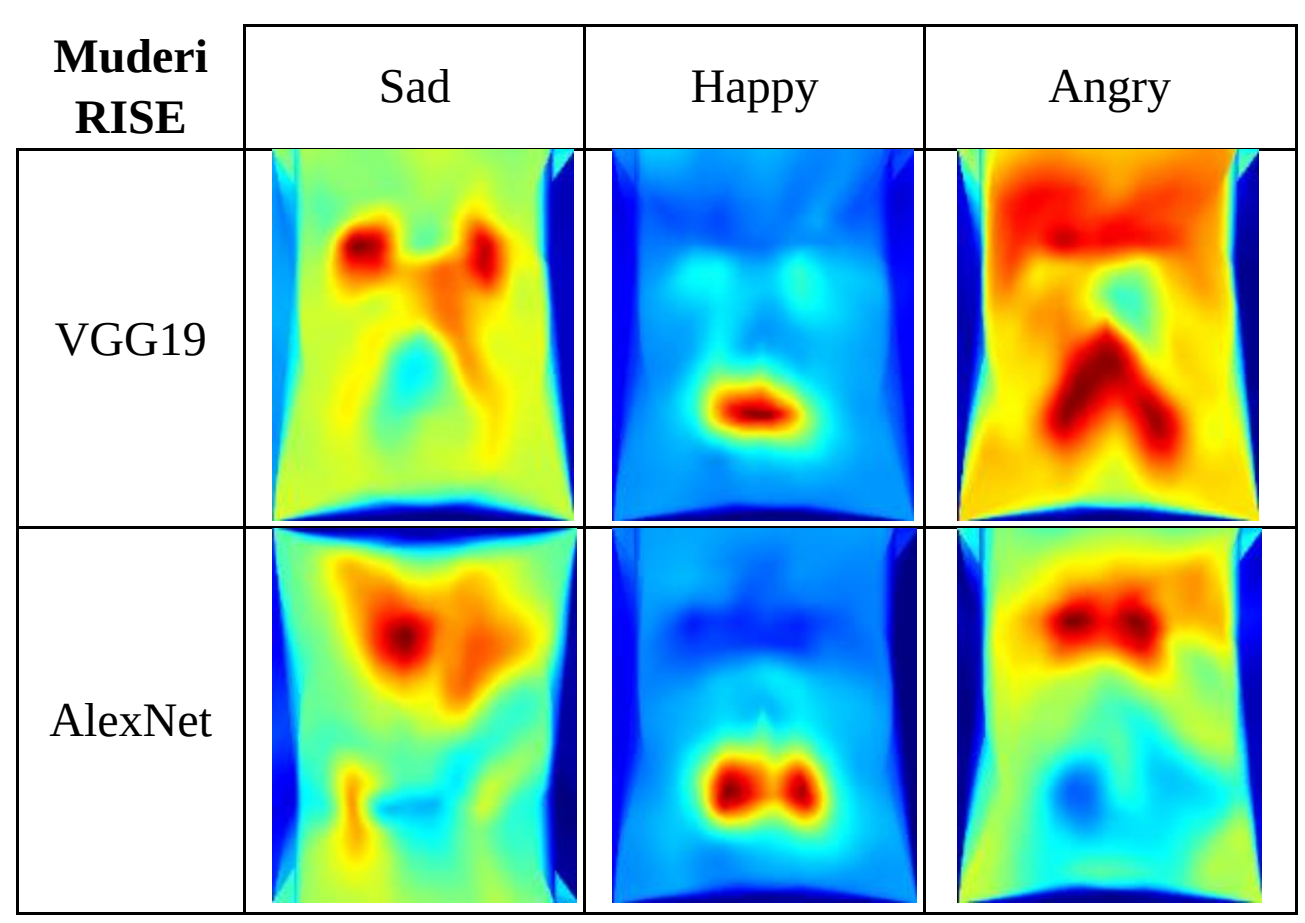

**Figure 9: Results of RISE on Muderi dataset using both VGG19 and AlexNet neural networks.**

#### *5.3.3 Differences between test datasets*

There are minor differences when changing the dataset between FE-Test and Muderi. Both LIME and RISE highlight more or less the same regions for both datasets for the same model, which indicates that the models are focusing on the same regions of the face for both datasets (Figures 6, 7, 8 and 9).

## 6 Discussion

From the obtained results, although the AlexNet and VGG19 networks focus on different regions of the face, they still obtain good results for the FE-Test dataset. Moreover, the LIME and RISE techniques usually coincide on which are the important regions for the two models, which increases the reliability of the explanations. Although the two models perform similarly for the FE-Test dataset, VGG19 is outperformed by the AlexNet on the Muderi dataset, which may be due to the pretraining of the VGG19, which AlexNet has not. However, the accuracy is still poor for the two models on the Muderi dataset, being under 50% in both cases. So, answering Q1: "Can neural networks trained with facial expressions predict the facial expressions of people with intellectual disabilities?". The results suggest that the models trained with people without intellectual disabilities cannot be used to recognize the facial expressions of people with them.

To answer Q2: "Are there differences between the facial expressions of people with intellectual disabilities and without any disabilities?" we can look at the heatmaps. It is clear that the models focus on the same regions for the two datasets (for people with and without intellectual disabilities), but still the accuracies obtained are very different. So, we can tell that there is a significant difference in how people of the Muderi dataset with intellectual disabilities manifest the three facial expressions, which leads the models to wrong predictions.

## 7 Conclusions

A facial expression recognition study for people with intellectual disabilities has been performed. We have carried out three experiments to answer the two initial questions: (Q1) Can neural networks trained with facial expressions predict the facial expressions of people with intellectual disabilities? (Q2) Are there differences between the facial expressions of people with intellectual disabilities and without any disabilities?

The obtained results show that the models trained on people without intellectual disabilities cannot be used to recognize the facial expressions of people with intellectual disabilities, at least for the Muderi dataset. On the other hand, in this case, the simplest model, AlexNet got better results than the more complex one, VGG19, although the results were still poor.

By using the XAI techniques, LIME and RISE, to explain the models' predictions, we have observed that the trained models, although focusing each on different regions for each expression, still focus on the same facial regions when changing the dataset. For this reason, there seem to be important differences between the facial expressions of people with and without intellectual disabilities.

In future work, we pretend to study more CNNs along with their explanations, and to add people with intellectual disabilities to the training dataset, to observe if a more specialized training leads to different explanations and better results. Finally, we intend to implement the system in a social robot and use it on people with intellectual disabilities.

## ACKNOWLEDGMENTS

This work has been supported by the Agencia Estatal de Investigación, Grant PID2019-104829RA-I00 funded by MCIN/ AEI/10.13039/501100011033, EXPLainable Artificial Intelligence systems for health and well-beING (EXPLAINING).

In addition, we also acknowledge the importance of the FPU scholarship from the Ministry of European Funds, University and Culture of the Government of the Balearic Islands.